\documentclass[jkps,preprint,fleqn,showpacs,showkeys]{revtex4}
\usepackage[pdftex]{graphicx}
\usepackage{amssymb}
\usepackage{amsmath}
\usepackage{bm}
\begin{document}
\setcounter{page}{0}
\title[]{Effect of the Collimator Angle on Dosimetric Verification of the Volumetric Modulated Arc Therapy}

\author{Yong Ho \surname{Kim}}
\author{Ha Ryung \surname{Park}}
\author{Won Taek \surname{Kim}}
\author{Dong Won \surname{Kim}}
\author{Yongkan \surname{Ki}}
\author{Juhye \surname{Lee}}
\author{Jinsuk \surname{Bae}}
\author{Dahl \surname{Park}}
\email{dpark411@gmail.com}
\thanks{Tel:+82-51-240-7924, Fax: +82-51-248-5747}
\affiliation{Department of Radiation Oncology, Pusan National University School of Medicine; Biomedical Research Institute, Pusan National University Hospital, Busan, 602-739}

\author{Hosang \surname{Jeon}}
\author{Ji Ho \surname{Nam}}
\affiliation{Department of Radiation Oncology, Pusan National University Yangsan Hospital, Yangsan, 626-770}


\begin{abstract}
Collimator angle is usually rotated when planning volumetric modulated arc therapy (VMAT) due to the leakage of radiation between multi-leaf collimator (MLC) leaves. We studied the effect of the collimator angles on the results of dosimetric verification of the VMAT plans for head and neck patients. We studied VMAT plans for 10 head and neck patients. We made 2 sets of VMAT plans for each patient. Each set was composed of 10 plans with collimator angles of 0, 5, 10, 15, 20, 25, 30, 35, 40, 45 degrees. Plans in the first set were optimized individually and plans in the second set shared the 30 degree collimator angle optimization. Two sets of plans were verified using the 2-dimensional ion chamber array MatriXX (IBA Dosimetry, Germany). The comparison between the calculation and measurements were made by the $\gamma$-index analysis. The $\gamma$-index (2\%/2 mm) and (3\%/3 mm) passing rates had negative correlations with the collimator angle. Maximum difference between $\gamma$-index (3\%/3 mm) passing rates of different collimator angles for each patient ranged from 1.46\% to 5.60\% with an average of 3.67\%. There were significant differences (maximum 5.6\%) in the passing rates of different collimator angles. The results suggested that the accuracy of the delivered dose depends on the collimator angle. These findings are informative when choosing a collimator angle in VMAT plans.
\end{abstract}



\maketitle

\section{INTRODUCTION}
Intensity modulated radiation therapy (IMRT) is usually used for head and neck cancer patients because it delivers highly conformal radiation doses to the target with reduction of toxicity to normal organs, as compared with conventional radiation therapy techniques \cite{1,2,3,4,5,6}. Volumetric modulated arc therapy (VMAT) is a novel IMRT technique. VMAT has less MU, less treatment time, high quality planning and more efficiency than static gantry angle IMRT \cite{1,2,3,4,5,6}. During VMAT the linear accelerator (LINAC) control system changes the dose rate and the multi leaf collimator (MLC) positions while gantry is rotating around the patient. 
Collimator angle is usually rotated in the plans of VMAT to reduce radiation leakage between MLC leaves. At a zero angle, the leakage between MLC leaves accumulates during the gantry rotation and the summed leakage results in unwanted dose distributions, which can not be controlled by optimization. At different collimator angles, the unwanted doses can be controlled by dose constraints in the optimization procedure so that  we can reduce the unwanted doses. The optimal collimator angle for VMAT plan is thus required to be determined. There are several factors for consideration in the choice of the collimator angle of the VMAT plan. Among them we concentrated on the accuracy of the VMAT delivery. We studied the effect of the collimator angle on the results of dosimetric verifications of the VMAT plan for nasopharyngeal cancer (NPC).

\section{Materials and Methods}
Ten patients with late-stage nasopharyngeal cancer were treated with concurrent chemo radiation therapy (CCRT). Eight patients had Stage III disease and 2 patients had Stage IV disease according to American Joint Committee on Cancer staging system 7.  Nine patients were male and 1 patient was female. One  radiation oncologist delineated radiation targets and organs at risk (OARs). The clinical target volume (CTV) included the primary nasopharyngeal tumor, neck nodal region and subclinical disease. Considering the setup uncertainty, margins ranging from 3-10 mm were added to each CTV to create a planning target volume (PTV). Reduced-field techniques were used for delivery of the 66-70 Gy total dose. The treatment plan course for each patient consisted of several sub-plans. In this study, we selected the first plan with prescribed doses of 50-60 Gy in 25-30 fractions to study the effect of the collimator angles on dosimetric verifications of the VMAT. The radiation treatment planning system Eclipse V10.0.42 (Varian Medical Systems, USA) was used to generate VMAT plans. The VMAT (RapidArc: Varian) plans were generated for Clinac IX linear accelerator using 6 MV photons. The Clinac IX is equipped with a Millennium 120 MLC that has spatial resolution of 5 mm at the isocenter for the central 20 cm region and of 10 mm in the outer 2$\times$10 cm region. The maximum MLC leaf speed is 2.5 cm/s and leaf transmission is 1.8\%.
Dosimetric leaf gap of the MLC was measured using the procedure recommended by Varian Medical Systems. The value of the dosimetric leaf gap was 1.427 mm for 6 MV photons.
For volume dose calculation, grid size of 2.5 mm, inhomogeneiy correction, the Anisotropic Analytical Algorithm (AAA) V10.0.28 and the Progressive Resolution Optimizer (PRO) V10.0.28 were used in all plans. VMAT plans for NPC patients were composed of 2 coplanar full arcs in 181-179 degree clockwise and 179-181 degree counterclockwise directions. The 2 full-arc delivery was expected to achieve better target coverage and conformity than the single arc \cite{7}. We generated 10 VMAT plans (plan set A) with different collimator angles for each patient. Ten collimator angles for the first arc were 0, 5, 10, 15, 20, 25, 30, 35, 40 and 45 degrees. For the second arc, the collimator angle was selected explementary to the collimator angle of the first arc in the same plan, i.e., the 2 collimator angles added up to 360 degree. The average field size of VMAT plans was 22 $\times$ 22 $\rm{cm^2}$. We used the same dose constraints for all the 10 VMAT plans and optimization was conducted for each plan. The maximum dose rate was 600 MU/min. The target coverage was aimed to achieve a 100\% volume covered by 95\% of prescribed dose. 
Optimization of each plan resulted in different fluences and different MLC motions for each plan.
Therefore we had 2 variables, i.e., the collimator angle and MLC motions.
To simplify the analysis we generated another set of 10 plans (plan set B) with the same MLC motions and different collimator angles for each patient.
The MLC motions were those of the plan with 30 degree collimator angle.
The plans in this set had different dose distributions and usually can not be used for treatment purposes excepting the plan with a 30 degree collimator angle.
\begin{figure}
\includegraphics[width=10.0cm]{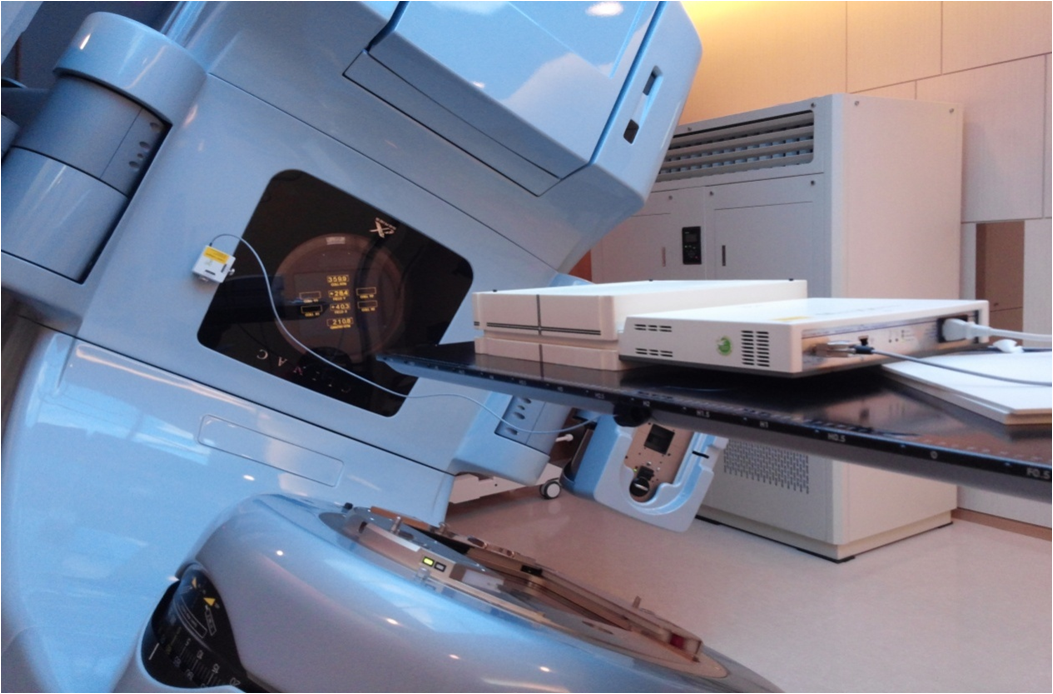}
\caption{Measurement devices (MatriXX, MultiCube and gantry angle sensor) set-up on the treatment couch for patient specific quality assurances of VMAT plans.
}\label{fig1}
\end{figure}

\begin{figure}
\includegraphics[width=10.0cm]{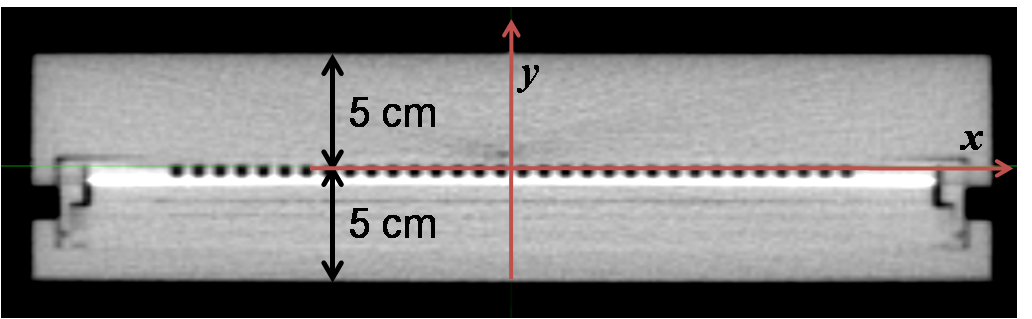}
\caption{Axial CT images of the MatriXX placed between solid water phantoms.
}\label{fig2}
\end{figure}
We performed patient specific quality assurances (QA) of 2 sets of 10 VMAT plans for each patient.
The measurements were made by the 2-dimensional ion chamber array MatriXX (IBA Dosimetry, Germany) \cite{8}.
The MatriXX has 1020 pixel ion chambers arranged in a 32$\times$32 matrix covering 24.4$\times$24.4 $\rm{cm^2}$.
Each ion chamber has the following dimensions: 4.5 mm in diameter, 5 mm in height and a sensitive volume of 0.08 $\rm{cm^3}$.
The distance between chambers is 7.619 mm.
The MatriXX has an intrinsic buildup and backscatter thicknesses of 0.3 mm and 3.5 mm, respectively.
The MatriXX was placed between solid water phantoms MultiCube (IBA Dosimetry, Germany) (Figure \ref{fig1}) so that thickness of total buildup and backscatter was 5 cm (Figure \ref{fig2}).
The source to surface distance was 95 cm with the measurement plane of the MatriXX at the isocenter of the LINAC. 
Measurement was done for each arc in the plan; therefore, we conducted 40 measurements for each patient and the total number of measurements was 400.
The angular dependence of the MatriXX was corrected after the measurements using the gantry angle sensor \cite{9} (IBA Dosimetry, Germany).
The comparison between the calculations and the measurements were made by $\gamma$-index (2\%/2 mm, 3\%/3 mm) analysis \cite{10,11,12} using OmniPro IMRT V1.7b (IBA Dosimetry, Germany). The $\gamma$-index was calculated only for the regions that have dose values above 10\% \cite{13} in the measured area.

\section{Results and Discussion}
\begin{figure}
\includegraphics[width=15.0cm]{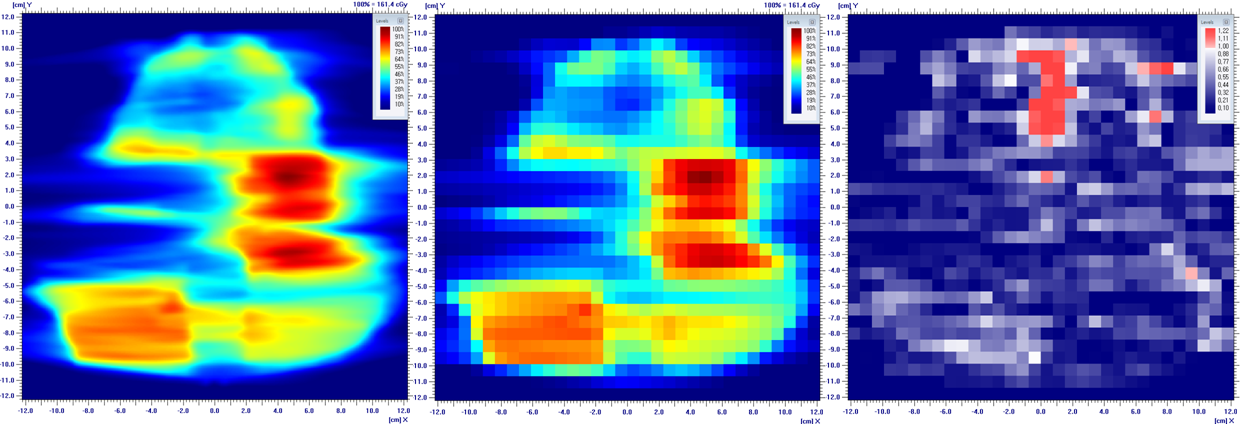}
\caption{Results for 1 patient plan in the plan set A for collimator angle 5. The first figure is the 2-dimensional dose distribution calculated by the Eclipse treatment planning system. The second one is the dose distribution measured by the MatriXX detector. The last one is the $\gamma$-index (3\%/3 mm) distributions. In the $\gamma$-index distributions red color indicates the region where the 3\%/3 mm criteria failed.
}\label{fig3-1}
\end{figure}

\begin{figure}
\includegraphics[width=15.0cm]{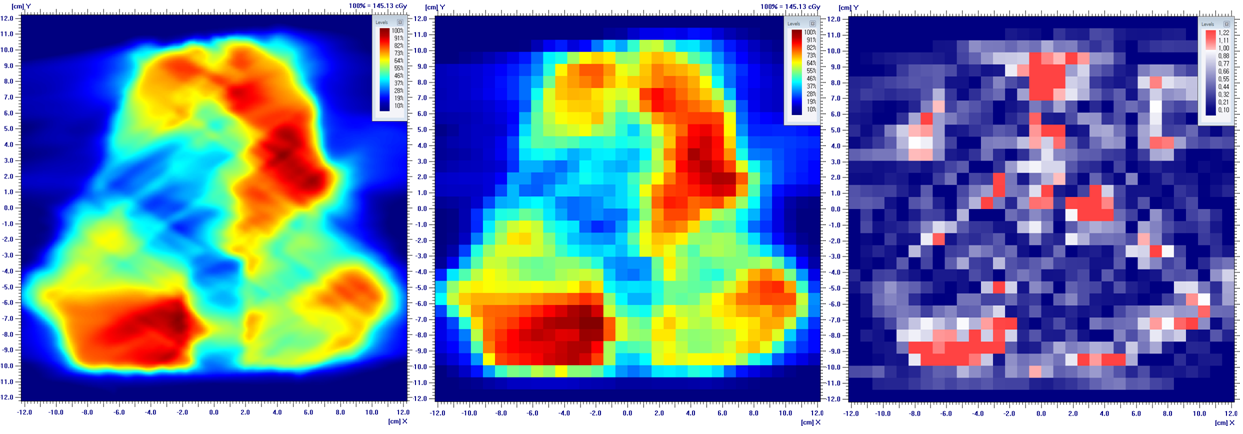}
\caption{Results for 1 patient (same patient in Figure \ref{fig3-1}) plan in the plan set A for collimator angle 35. The first figure is the 2-dimensional dose distribution calculated by the Eclipse treatment planning system. The second one is the dose distribution measured by the MatriXX detector. The last one is the $\gamma$-index (3\%/3 mm) distributions. In the $\gamma$-index distributions red color indicates the region where the 3\%/3 mm criteria failed.
}\label{fig3-2}
\end{figure}

\begin{figure}
\includegraphics[width=10.0cm]{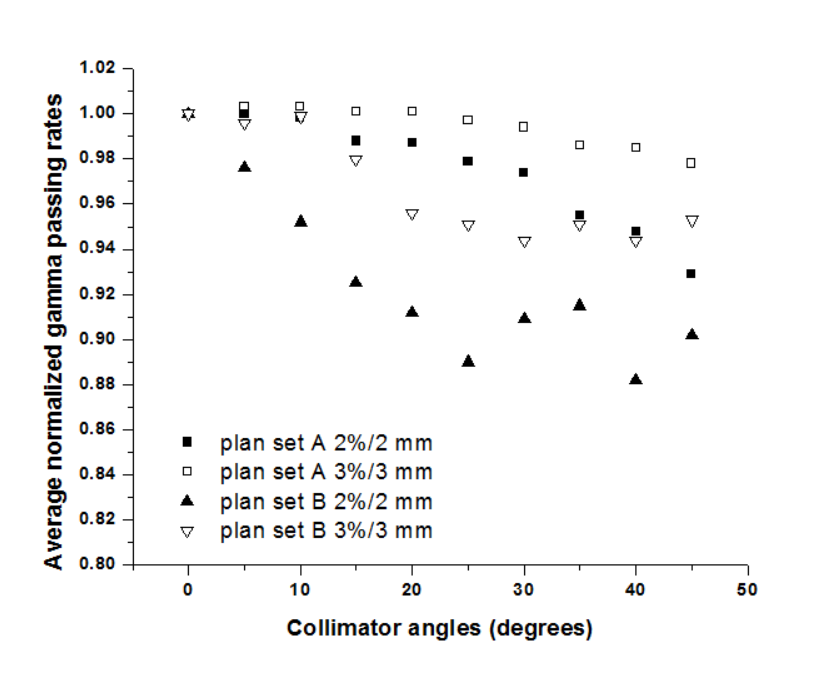}
\caption{Average normalization values for the $\gamma$-index passing rates (2\%/ 2 mm) of patient specific delivery QAs as a function of the collimator angle. 
}\label{fig4}
\end{figure}

Average $\gamma$-index passing rates of patient specific QAs were given in Table \ref{table1}.
The results were averaged over the 2 arcs and 10 patients.
Because the 2 arcs in each VMAT plan rotated almost 360 degrees and the measurement set-up is mirror symmetric about the measurement plane ($y=0$ plane in Figure \ref{fig2}) of the MatriXX detector and a vertical plane passing through the isocenter ($x=0$ plane in Figure \ref{fig2}) the arc with collimator angle $\theta$ is symmetric to the arc with collimator angle $360-\theta$.
Therefore we regarded the collimator angle of the second arc, which was equal to 360 minus the collimator angle of the first arc, as the collimator angle of the first arc in the analysis.
\begin{table}
\caption{$\gamma$-index passing rates of the patient specific QAs as a function of the collimator angle 
}
\begin{ruledtabular}
\begin{tabular}{ccccc}
Collimator & \multicolumn{2}{c}{Plan set A} & \multicolumn{2}{c}{Plan set B} \\
angle & 2\%/2 mm & 3\%/3 mm & 2\%/2 mm & 3\%/3 mm \\
\colrule
0 & 91.87$\pm$3.45\% & 97.32$\pm$1.74\% & 87.95$\pm$5.35\% & 97.07$\pm$2.36\% \\
5 & 91.86$\pm$3.39\% & 97.59$\pm$1.77\% & 87.37$\pm$4.66\% & 96.81$\pm$2.02\% \\
10 & 91.65$\pm$3.16\% & 97.63$\pm$1.41\% & 87.00$\pm$4.67\% & 97.05$\pm$1.61\% \\
15 & 90.75$\pm$3.17\% & 97.46$\pm$1.57\% & 86.82$\pm$5.20\% & 96.60$\pm$2.15\% \\
20 & 90.64$\pm$3.69\% & 97.41$\pm$1.45\% & 85.86$\pm$5.20\% & 95.86$\pm$2.66\% \\
25 & 89.93$\pm$3.70\% & 97.05$\pm$1.74\% & 83.97$\pm$5.74\% & 94.70$\pm$2.84\%  \\
30 & 89.52$\pm$2.92\% & 96.76$\pm$1.65\% & 83.45$\pm$6.35\% & 94.13$\pm$3.11\% \\
35 & 87.73$\pm$4.39\% & 95.94$\pm$1.91\% & 84.00$\pm$6.21\% & 94.53$\pm$3.16\% \\
40 & 87.06$\pm$4.19\% & 95.87$\pm$1.49\% & 84.09$\pm$6.35\% & 94.53$\pm$3.35\% \\
45 & 85.37$\pm$5.18\% & 95.19$\pm$1.76\% & 83.95$\pm$6.41\% & 94.29$\pm$3.44\% \\
\end{tabular}
\end{ruledtabular}
\label{table1}
\end{table}
Maximum difference between $\gamma$-index (2\%/2 mm) passing rates of plans in plan set A for each patient ranged from 2.83\% to 14.32\% and the average value was 8.44$\pm$4.24\%.
Using the 3\%/3 mm criteria the maximum difference ranged from 1.46\% to 5.60\% and the average value was 3.67$\pm$1.29\%.
Maximum difference between $\gamma$-index (2\%/2 mm) passing rates of plans in plan set B for each patient ranged from 3.71\% to 10.44\% and the average value was 7.97$\pm$2.17\%.
Using the 3\%/3 mm criteria the maximum difference ranged from 1.46\% to 7.23\% and the average value was 4.69$\pm$2.51\%.
2-dimensional dose distributions calculated by the Eclipse treatment planning system, dose distributions measured by the MatriXX detector and $\gamma$-index (3\%/3 mm) distributions of 1 patient plans in the plan set A for collimator angle 5 and 35 degree were shown in Figure \ref{fig3-1} and \ref{fig3-2}, respectively.
The passing rate for the 35 degree collimator angle was less than the passing rate for the 5 degree collimator angle.
The increase in collimator angle resulted in decreased $\gamma$-index passing rates, as shown in Figure \ref{fig4}.
In the Figure, passing rates were normalized to the value of 0 degree.
Black and white squares indicated $\gamma$-index (2\%/2 mm) and (3\%/3 mm) passing rates, respectively, averaged over plan set A.
Black and white triangles indicated $\gamma$-index (2\%/2 mm) and (3\%/3 mm) passing rates, respectively, averaged over plan set B.
There were statistically significant negative correlations between the collimator angle and the $\gamma$-index passing rates.
Pearson correlation coefficients for pair-wise ratings of the $\gamma$-index (2\%/2 mm) and (3\%/3 mm) passing rates of plans in the plan set A and B were -0.524 and -0.412, respectively with p-values $<$ 0.001.
For accuracy of VMAT a smaller collimator angle is better, and for MLC leakage a larger collimator angle is better, we were thus required to make a compromise.
Based on this study, in our hospital the collimator angles of the VMAT plans for head and neck patients range between 15-25 degrees because the average $\gamma$-index passing rates were above or near to 90\% for the 2\%/2 mm criteria and 97\% for the 3\%/3 mm criteria, as shown in the results of the passing rates for the plan set A (Table \ref{table1}). In other hospitals these results can be somewhat different because they have different VMAT delivery systems and diffterent VMAT planning systems. We think that they can find optimal collimator angles by conducting the similar measurements described in this article.

Although not included in this article, we performed the patient specific QAs for other treatment sites with smaller field sizes that are $<$ 13 $\times$ 13 $\rm{cm^2}$.
Maximum difference of the passing rates for VMAT plans with various collimator angles was $<$ 1.5\%.
Collimator angle does not affect the accuracy of the VMAT delivery with small field sizes.

The accuracy of radiation delivery by the LINAC depends on geometrical accuracies such as gantry isocentricity, collimator isocentricity and MLC position. It was reported that leaf limiting velocity, MLC position and mechanical isocenter varied at different collimator and gantry angles \cite{14,15}. This may explain the $\gamma$-index passing rates dependence on the collimator angle. Further study is needed to investigate the origin of the collimator angle dependence of the accuracy of VMAT delivery.

The quality of the plan itself is another factor for consideration in the choice of the collimator angle of the VMAT plan.
Optimized dose distributions with the same dose constraints can vary according to the collimator angle of the VMAT plan.
Further study is needed to evaluate the quality of VMAT plans with different collimator angles.

\section{CONCLUSIONS}
We found that the results of the patient specific QAs for VMAT plans using the 2-dimensional ion chamber array MatriXX are dependent on the collimator angle of the VMAT plans. The $\gamma$-index (2\%/2 mm) and (3\%/3 mm) passing rates were negatively correlated with the collimator angle.
We showed that collimator angles of the VMAT plans for head and neck cancer patients range between 15-25 degrees resulting in the average $\gamma$-index passing rates above or near to 90\% for the 2\%/2 mm criteria and 97\% for the 3\%/3 mm criteria.

\end{document}